 \documentclass [ashowpacs,preprintnumbers,amsmath,amssymb]{revtex4}
\usepackage{hyperref}

\usepackage{graphicx}

\usepackage{dcolumn}
\usepackage{amssymb}
\newcommand{\be}{\begin{eqnarray}}
\newcommand{\ee}{\end{eqnarray}}

\newcommand{\np}[1]{Nucl. Phys. {\bf #1}}

\begin{document}

\title {Dynamical effects in proton breakup from exotic nuclei}

\author{ Ravinder Kumar$^{a,b}$ and  Angela  Bonaccorso$^{a}$ \\ \small
  $^{a}$INFN, Sez. di Pisa and $^{b}$Dipartimento di Fisica, Universit\`a di Pisa,\\\small Largo Pontecorvo 3, 56127
Pisa, Italy.\\ \small }

\begin{abstract}
 We study dynamical effects in proton breakup  from a weakly bound state in an exotic nucleus on a heavy target. The Coulomb interactions between the proton and the core and  the proton and the target are treated to all orders, including also the full multipole expansion of the Coulomb potential. The dynamics of proton nuclear and Coulomb breakup is compared to that of an {\it equivalent} neutron of larger binding energy in order to elucidate the differences with the well understood neutron  breakup mechanism. A number of experimentally measurable observables such as parallel momentum distributions, proton  angular distributions and total breakup cross sections are calculated. With respect to nuclear breakup it is found that a proton behaves exactly as a neutron of larger binding energy. The extra "effective energy" is due to the combined core-target Coulomb barrier. In Coulomb breakup we distinguish the effect of the core-target Coulomb potential (called recoil effect), with respect to which the proton behaves again as a more bound neutron, from the direct proton-target Coulomb potential. The latter gives cross sections about an order of magnitude larger than the recoil term. The two effects give rise to complicated interferences in the parallel momentum distributions. They are instead easily separable in the proton angular distributions which are therefore suggested as a very useful observable for future experimental studies.
 \end{abstract}

\maketitle

{\bf Pacs} {21.10.Jx, 24.10.-i, 25.60.Gc, 27.30.+t}
\section{Introduction}

In a recent paper Liang et al.\cite{lia} studied experimentally dynamical effects in the Coulomb breakup of $^{17}$F. The motivation was that
 Coulomb
dissociation, being  the inverse of the radiative
capture reaction,  is a useful technique for studying
stellar nucleosynthesis involving short-lived nuclei where direct
measurements are difficult \cite{[1]}. The authors argued that with an increasing
number of radioactive isotope beams available for studying capture
reactions that occur in stellar environments, such as those in the
rp-process, more measurements  will be performed by
Coulomb dissociation because of the short lifetimes that make direct
measurements impractical. In order to extract reliable information
on radiative capture the reaction mechanism
in Coulomb dissociation has to be understood in detail. Furthermore Liang et al.\cite{lia} stressed the importance of the
dynamic polarization effect which they interpret as a displacement of  the valence proton 
behind the nuclear core and a subsequent shielding from the target. This effect
manifests itself as a reduction of breakup probability compared
to first-order perturbation theory predictions.
Coulomb dissociation experiments \cite{[3}-\cite{5]} have been studied with different approaches, among which first-order perturbation theory is indeed often used with final plane wave functions. While Coulomb dissociation
of loosely bound neutron-rich nuclei has been studied
extensively and is fairly well understood \cite{[10,11]}, this is not true
for proton-rich nuclei in which the loosely bound valence protons
actively participate in the reaction. It has also been  suggested that the inclusion
of higher-order corrections is required \cite{[12,13]}. Essentially exact
calculations of realistic three-body models of break-up based on the
Faddeev equations including the Coulomb potential exist for few body nuclei \cite{14}. These methods have  also been applied to neutron break up from medium mass exotic nuclei using cluster models \cite{15} and it is expected that they should perform equally well for proton breakup. 

Besides the well known astrophysical implications proton rich nuclei present a number of unusual features such as the two proton radioactivity or beta-delayed proton emission which make them very appealing to study and to compare to the neutron rich nuclei. An account of the richness of the "Physics of the proton rich side of the nuclear chart" can be found in Ref.\cite{euri}.

In Ref.\cite{lia} $^{17}$F + $^{58}$Ni and  $^{17}$F +
$^{208}$Pb breakup angular distributions measured at 10A.MeV were discussed and compared to first-order perturbation theory prediction for Coulomb breakup and to fully dynamical calculations. At the end of their paper the authors used first-order perturbation theory to calculate proton breakup via an increase of the effective binding energy of the valence proton
 treated as if it were a neutron, according to a model that was proposed some time ago by some of us \cite{ang04}. Liang et al. \cite{lia} found that our idea did indeed work and the angular distributions could successfully be reproduced using an increase in the binding energy of 1.2 MeV, contrary to our model suggestion of 3.2 MeV. 

We were intrigued and interested by such a result: first we were  happy to notice that our very schematic model would work in comparison with real data; second we decided that it was worth studying if and how the model could be made more quantitatively  reliable. This paper reports  results of such a study for some observables such as parallel momentum distribution and   proton  angular distributions following breakup. A forthcoming paper will discuss  core angular distributions.

In the next section we recall the model for nuclear and Coulomb breakup for a valence nucleon
treated to all orders introduced in \cite{nois2}, that we developed successively in Ref.\cite{nois} to treat protons.  Our
starting point is, as in all our previous works,  a first order perturbation theory
amplitude such as that given by Eq.(1) of  \cite{luigi}. In  that paper the
amplitude was used for transfer to bound states, thus the
neutron-target potential acted only to first order. On the other
hand when the same formula is used for transitions to final unbound states \cite{me}, in
order to describe breakup, the final state is a proper scattering
state with respect to the target and thus all re-scattering terms
are included \cite{mar02,mar03}. We then compare in Sec.III results of the full calculation to those of perturbation theory. The amount and characteristics of the breakup due to the core-target  and to the proton-target Coulomb repulsions  are then studied separately.  We do that at fixed impact parameter values in order to understand the reduction of breakup probability for protons as compared to that of neutrons of the same binding.  This is because our model predicted  that the effective binding energy would depend on the distance between the two nuclei, being due to the mutual projectile-target Coulomb barrier. The impact parameter dependence enters also in the model of    Ref. \cite{nois} where we showed that the Coulomb breakup could be treated by time dependent perturbation theory at large impact parameters, while it could be described by an all order eikonal model at small impact parameters. Sections III contains results for $^{8}$B, while Section IV is devoted to $^{17}$F. In Section V the proton and neutron angular distributions due to Coulomb breakup are shortly described. Finally our conclusions are given in Sec.VI.

  \section{Formalism} \label{teo}
\begin{figure}[h]
\center
\includegraphics[scale=1,width=8cm,angle=-90]{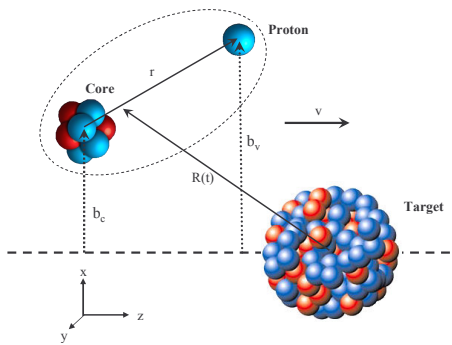}
\caption{(Color online). Coordinate system.}
\label{pall}
\end{figure}
The Coulomb potential responsible for proton breakup is
\begin{eqnarray}
V(\vec{r},\vec{R})= \frac{V_c}{|\vec{R}-\beta_1 \vec{r}|}+\frac{V_v}{|\vec{R}+\beta_2 \vec{r}|}-\frac{V_0}{R} \label{eq1}\end{eqnarray}
 where $V_c=Z_cZ_te^2$, $V_v=Z_vZ_te^2$ and $V_0=(Z_v+Z_c)Z_te^2$. $\beta_1$ and $\beta_2$ are the mass ratios of proton and core, respectively, to that of the projectile. Z$_c$, Z$_t$ and Z$_v$  are the core, target and proton charge respectively. The coordinate system used in this paper is shown in Fig. \ref{pall}. Following \cite{nois,nois2}, the perturbation theory phase is defined  as $\chi_{pert}=\frac{1}{\hbar}\int dt e^{i\omega t} V(\vec{r},t)$, which calculated explicitly gives 
 
 \begin{eqnarray} \label{fasipr}
\chi^p&=&\frac{2}{\hbar v} (V_c e^{i\beta_1 \omega z /v}K_0(\omega b_c/v) -V_0 K_0(\omega R_\perp/v) \nonumber  \\ &&+V_v e^{-i\beta_2 \omega z /v}K_0(\omega b_v/v)  )
\end{eqnarray}
with $b_v=b_c+r_{\perp}$, $\omega=\left(  {{\varepsilon  _f}}-\varepsilon_{0}\right)/\hbar
$ and $\varepsilon_0$ is the neutron initial bound state energy  while
${{\varepsilon  _f}}$ is the  final neutron-core continuum energy.
 Since $V_0=V_c+V_v$, Eq. (\ref{fasipr}) can be written as
\begin{eqnarray} \label{pff}
\chi^p=\chi(\beta_1,V_c)+\chi(-\beta_2,V_v)
\end{eqnarray}
 where each term is given by 
\begin{eqnarray} \label{fiqui}
\chi(\beta,V)=\frac{2 V}{\hbar v}\left(e^{i\beta \omega z /v}K_0(\omega b/v) -K_0(\omega R_\perp/v)\right),
\end{eqnarray}
$\chi(\beta_1,V_c)$  describes the recoil of the core whereas $\chi(-\beta_2,V_v)$ accounts for the direct proton-target Coulomb interaction. The latter vanishes in the case of the neutron.

 The expansion of Eq. (\ref{fiqui}) to first order in $\vec{r}$ yields the dipole approximation to the phase:
\begin{eqnarray}\label{dipp}
\chi^p &\simeq& \frac{2 (\beta_1V_c-\beta_2V_v)}{\hbar v}(K_0(\omega R_\perp/v)\frac{i\omega z}{v} \nonumber \\ &&+K_1(\omega R_\perp/v)\frac{\vec{R}_\perp \cdot \vec{r}}{R_\perp}\frac{\omega}{v}),
\label{chip}\end{eqnarray}
 The  constant factor is now $(V_c\beta_1-V_v\beta_2)$ while for a neutron would be $V_0\beta_1$  \cite{nois}.
 
We recall here that in Ref.\cite{ang04} it was suggested that proton breakup could be treated as neutron breakup and  the Coulomb potential of Eq.(\ref{eq1}) could be approximated by a dipole expansion if the neutron was given an effective binding energy which would take into account the combined effect of the core and target Coulomb barriers.
% \begin{equation} {\tilde\varepsilon}_i=  \varepsilon_i-\Delta=\varepsilon_i-{Z_pe^2\over R_i}-Z_te^2\left ({1\over 2}\left (  {1\over{|d+R_i|} }+ {1\over{|d-R_i|} }\right )-{1\over d}\right )\label {1a}\end{equation}
  %\begin{equation}
%V(\vec{r},\vec{R})\approx V_{c0} + K\frac{\vec{r}\cdot \vec{R}}{|\vec{R}|^3}
%\end{equation}
%the constants 
%\begin{equation}
%V_{c0}= \frac{1}{2}\left(V_{coul}(R_i, R) + V_{coul}(-R_i, R) \right)
%\end{equation}
%and 
%\begin{equation}
%K\frac{R_i}{R^2}= \frac{1}{2}\left(V_{coul}(R_i, R) - V_{coul}(-R_i, R) \right)
%\end{equation}
%are defined such that in the limit when $R$ is very large the dipole expansion is good and we have $V_{c0}\approx 0$ and
%$K\approx \beta_2 Z_t Z_v e^2 -\beta_1 Z_t Z_c e^2$, which consistently would lead to the phase Eq.(\ref{chip}). 
As we mention in the Introduction, this model has been applied in Ref.\cite{lia}. Liang et al.  used first-order perturbation theory to calculate proton breakup via an increase of the effective binding energy of the valence proton
 treated as if it were a neutron. The authors fitted their the angular distributions  using an increase in the binding energy of 1.2 MeV. We shall show in the following that accurate calculations of proton vs neutron breakup suggest indeed that in certain cases proton breakup can be understood in terms of a more bound neutron breakup. However we believe now that a better estimate of the effective energy is given by

 \begin{equation} {\tilde\varepsilon}_i=  \varepsilon_i-\Delta=\varepsilon_i-{Z_pe^2\over  R_i}-Z_te^2\left ({1\over 2}\left (  {1\over{|d+\beta_2 R_i|} }+ {1\over{|d- \beta_1 R_i|} }\right )
-{1\over d}\right ),\label {1abis}\end{equation}
which through the factors $\beta_1$ and $\beta_2$ is  consistent with Eq.(\ref{eq1}).  $R_i$  the position of the projectile top of the Coulomb barrier, $d$ is the distance between the center of the two nuclei for which the top of the two Coulomb barriers of projectile and target coincide.

Our expression for the differential cross-section is
\begin{eqnarray}
\frac{d \sigma}{d \vec{k}}=\frac{1}{8\pi^3}\int d \vec{b}_c |S_{ct}(b_c)|^2 |g^{rec}+g^{dir}+g^{nuc}|^2.
\label{cross}\end{eqnarray}
where $|S_{ct}(b_c)|^2$ is the core-target elastic scattering probability. 

The probability amplitude in Eq.(\ref{cross}) has been written as the sum of three pieces:
the recoil term,
\begin{eqnarray}
g^{rec}&=&\int d\vec{r} e^{-i \vec{k} \cdot \vec{r}} \phi_i(\vec{r}) ( e^{i \frac{2 V_c}{\hbar v}\log{\frac{b_c}{R_\perp}}} -1 -i\frac{2 V_c}{\hbar v}\log{\frac{b_c}{R_\perp}} \nonumber \\ && +i\chi(\beta_1,V_c) ),
\label{grec}\end{eqnarray}
 obtained in the sudden limit according to  \cite{nois,nois2}  in order to include all orders in the final state Coulomb interaction of the core  with the target. Similarly, the second term in our probability amplitude is the direct proton Coulomb interaction. It has the same form as Eq.(\ref{grec}) but for the substitution $V_c\to V_v$, $b_c \to b_v$ and $\beta_1 \to -\beta_2$. Both the direct and recoil term contain a regularization of the first order term divergency consisting in substituting it with the corresponding first  order perturbation theory term. Such regularization method was first proposed in Ref.\cite{mar03} and used also in Refs.\cite{suz,suz1}.\\
 Finally, the nuclear part is 
\begin{eqnarray}
g^{nuc}=\int d\vec{r} e^{-i \vec{k} \cdot \vec{r}} \phi_i(\vec{r}) \left( e^{i \chi_{nt}(b_v)} -1 \right).
\end{eqnarray}
 
 % A number of papers have addressed the problem of asymmetry in the core parallel momentum distribution after proton knockout \cite{davids98,davids01,davids03,esben96,esben00,esben95,tt}. The fact that this asymmetry comes from high order terms can be directly extracted from our formalism. 
 
 If the Coulomb part of the amplitude $g^{Cou}=g^{rec}+g^{dir}$ is simply expanded to first order in $\chi$  and it is written, in terms of the one-dimensional Fourier transform in $z-$direction $\hat{\phi}_i$, then making also the dipole approximation one gets:
% \begin{eqnarray}
%g^{Cou}&\simeq &\int d\vec{r}_\perp e^{i \vec{k}_\perp \cdot \vec{r}_\perp} \frac{2}{\hbar v}\left(V_c K_0(\omega b_c/v)\hat{\phi}_i(\vec{r}_\perp,k_z+\beta_1 \omega/v )\right. \nonumber \\ && -\left.V_0 K_0(\omega R_\perp/v)\hat{\phi}_i(\vec{r}_\perp,k_z) \right. \nonumber \\ &&
% +\left.V_v K_0(\omega b_v/v)\hat{\phi}_i(\vec{r}_\perp,k_z-\beta_2 \omega/v)\right)\label{11}
%\end{eqnarray}

\begin{eqnarray}
g^{Cou}&\simeq &\int d\vec{r}_\perp e^{-i \vec{k}_\perp \cdot \vec{r}_\perp}\frac{2 (\beta_1V_c-\beta_2V_v)}{\hbar v} \nonumber \\
&& \times\left(K_0(\omega R_\perp/v)\frac{\omega}{v}\frac{d}{d k_z}\hat{\phi}_i(\vec{r}_\perp,k_z)  \right.\nonumber \\
&&\left.+K_1(\omega R_\perp/v)\frac{\vec{R}_\perp \cdot \vec{r}}{R_\perp}\frac{\omega}{v}\hat{\phi}_i(\vec{r}_\perp,k_z)\right),
\label{dip}\end{eqnarray}
 leading to symmetric momentum distributions. 
  
 A number of experimental data for neutron breakup have been analyzed by separating high impact parameters such that perturbation theory would be valid. We shall try in the following to do the same for proton breakup and in  order to clarify the effects of the Coulomb interaction we will show separately results  for the Coulomb or nuclear  probability only. Furthermore we will show momentum probability distributions at fixed impact parameters, namely:
 
% First we study the effect of using neutron wave functions of larger binding energy, according to Eq.(\ref{1a}) instead than proton wave functions. In Fig.\ref{1} we show the proton wave function corresponding to the $p_{1/2}$ ground state together with neutron wave functions of binding energy $\varepsilon_i=0.7~{\rm and}~ 0.8 $ MeV

\begin{eqnarray}
\frac{dP(b_c)}{d {k_z}}= \frac{1}{8\pi^3}\int d \vec{k}_{\perp}|g(b_c)|^2.
\label{pdis}\end{eqnarray}
$|g(b_c)|^2$ will be the nuclear or Coulomb amplitude and in the latter case we will indicate in which approximation the calculation was done.

\section{Application to $^8$B}

\begin{figure}[h]
\center
\vskip 15pt
\includegraphics[scale=1,width=9cm,angle=0]{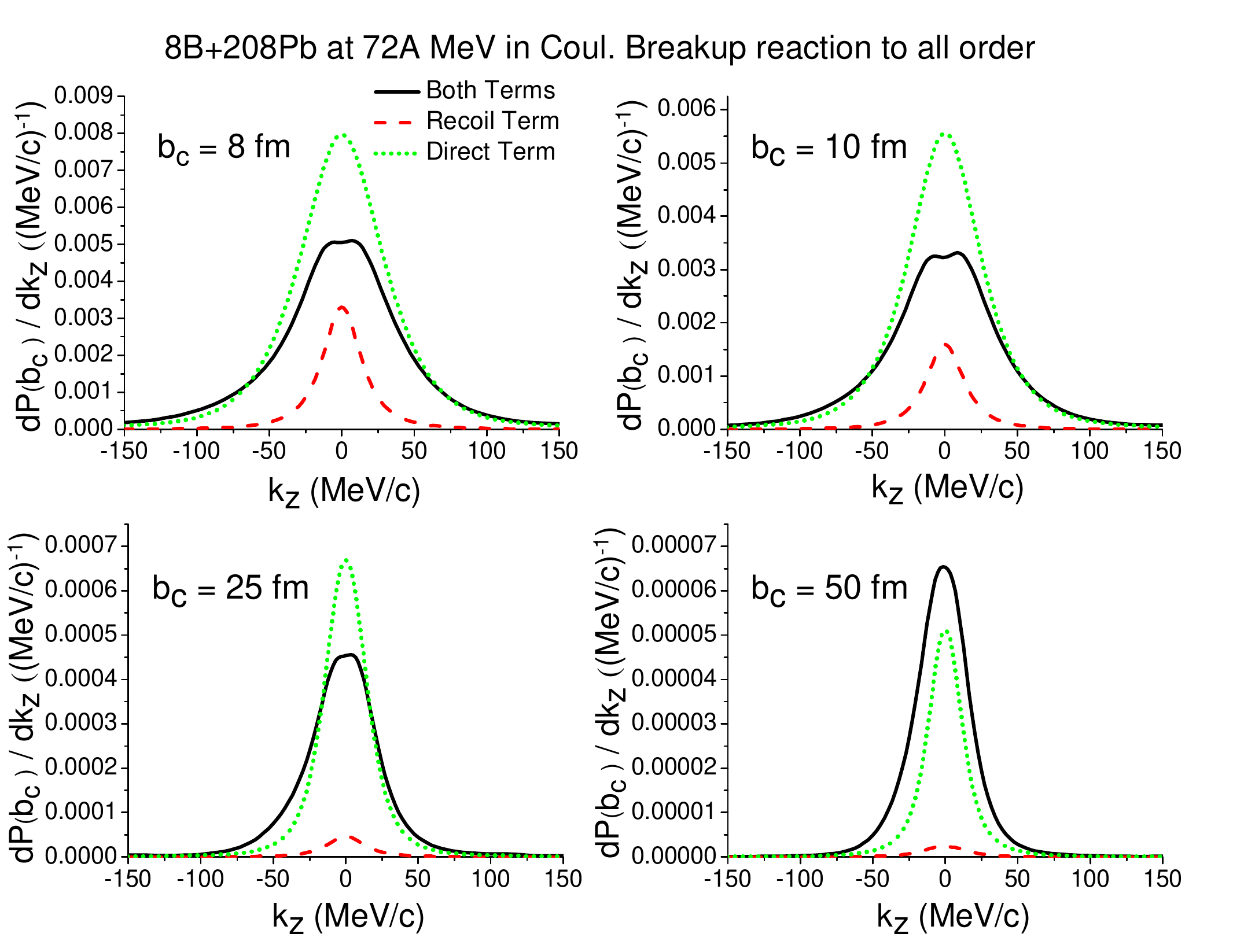}
\caption {(Color online). Recoil term, Eq. (\ref{grec}), red dashed line and  direct term, green dotted line. The black full line contains both.}
\label{f4}
 \end{figure}
\begin{table}[h]
\caption{Barrier radii, initial binding energies and effective energy parameters for a $^{208}$Pb target.}
\vskip.3in \begin{center}
\begin{tabular}{lccccccc}
\hline\
                     & $^{8}{\rm B}$&J$^\pi$& $^{17}{\rm F}$&J$^\pi$\\ \hline
R$_{i}$(fm) & 6.0 && 6.5&\\
$\varepsilon_i$(MeV) & -0.14 &1p$_{3/2}$    & -0.6 &1d$_{5/2}$ \\
$\varepsilon_i^*$(MeV) &-0.57 &1p$_{1/2}$&-0.1 & 2s$_{1/2}$ \\
$-\Delta$(MeV)&-0.4&&-1.2&&\\
$\tilde{\varepsilon}_i$(MeV) & -0.54 &1p$_{3/2}$    & -1.8 &1d$_{5/2}$ \\
$\tilde{\varepsilon}_i^*$(MeV) &-0.97 &1p$_{1/2}$&-1.3 & 2s$_{1/2}$ \\
\hline
\end{tabular}\end{center}
\end{table}

In our previous work in which the method presented here was introduced, we applied it to study a reaction at 936A.MeV, the energy was high enough to justify the eikonal approximation. Before starting the calculations presented in this and the following  section we have compared our results for the total breakup probabilities against the values provided by Ref.\cite{esb02}. We have found that for nuclear breakup our results agree with the dynamical calculations down to 10A.MeV if we substitute the impact parameter by the distance of closest approach of the corresponding classical trajectory. For Coulomb breakup our probabilities decrease with the impact parameter as those of Ref.\cite{esb02}
but they are about 50\% larger. This is due to the use of a straight line relative motion trajectory in our method. For nuclear breakup the substitution of the impact parameter with the distance of closest approach is enough to cure the method, because the nuclear form factors are different from zero only in  a small region around  the distance of closest approach where the classical trajectory is well approximated by a  straight line. On the other hand the same cannot be said of the Coulomb form factor. We have checked however that our method starts to give  reliable absolute probability values around 70-100A.MeV which agrees with what we found in our previous works on neutron breakup \cite{mar02,mar03}.

%\begin{figure}[h]\center\vskip 15pt\includegraphics[scale=0.6,width=7cm,angle=-90]{Fig5.pdf}\caption {Nuclear breakup terms: stripping, black full   line and diffraction, red dashed. Coulomb breakup to all oders, dotted green curve and Coulomb plus diffraction (including interference), blue dot-dashed line.}\label{f5}\end{figure}

\begin{figure}[h]\center\vskip 15pt\includegraphics[scale=0.9,width=8cm,angle=0]{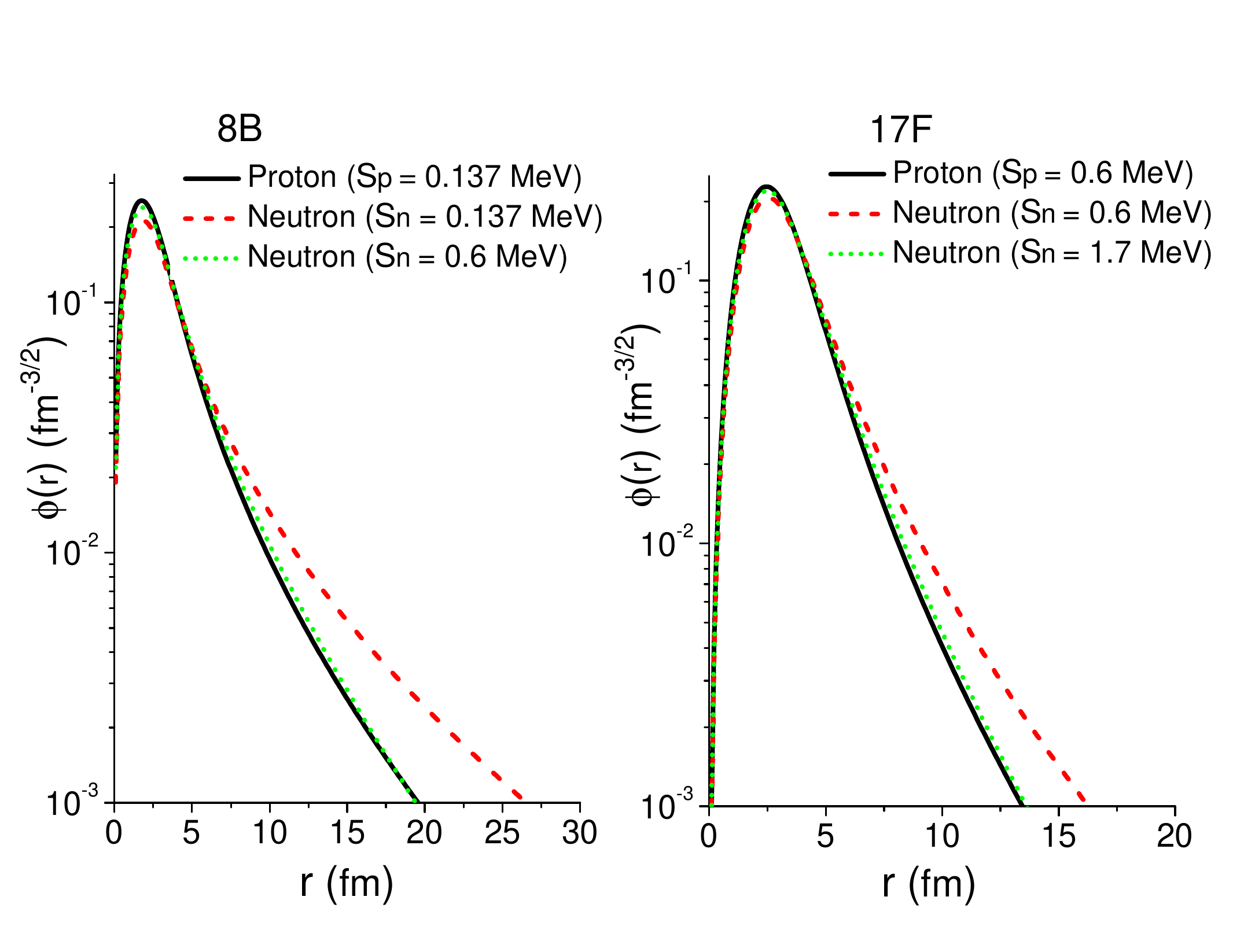}\caption {(Color online). Proton vs neutron  wave functions for a p$_{3/2}$ single particle state in a $^8$B and d$_{5/2}$ in $^{17}$F as indicated.}\label{f6}\end{figure}

%\begin{figure}\center\includegraphics[scale=0.6,width=7.cm,angle=-90]{1.pdf}\caption{Coulomb breakup probability distributions at fixed core-target impact parameter as a function of the parallel proton momentum for  proton-removal from the g.s of $^8$B against Pb at 72 A.MeV. Details are given in the text.}\label{f2}\end{figure}

%\begin{figure}
%\center
%\vskip 15pt
%\includegraphics[scale=0.2,width=7cm,angle=-90]{2.pdf}
%\caption {(Colour online) Proton momentum distribution after Coulomb breakup of  $^8$B against Pb at 936 A.MeV in both dipole and full-multipole approximations, for both ground and excited state. Spectroscopic factors are not included.}
%\label{gsi}
%\end{figure}
Thus the formalism described in the previous section has been applied to proton breakup of $^8$B and $^{17}$F against Pb  target at a beam energy of 72A.MeV which is a typical energy used in several laboratories and for which our results should be reliable.  
%For  $^{17}$F we have also done some exploratory calculations at 10A.MeV. 
The  projectiles are taken  as a two-body object. Radial wave functions have been obtained by numerical solution of the Schr\"odinger equation in Woods-Saxon potentials with depths adjusted to reproduce the experimental  separation energies (0.137MeV and 0.6MeV respectively). The radius parameter of these Woods-Saxon potentials has been taken as 1.3 fm and the diffuseness as 0.6 fm. 

% \begin{figure}\center\vskip 15pt\includegraphics[scale=0.6,width=7.cm,angle=-90]{Fig3.pdf}\caption {All order regularized Coulomb breakup calculation, black full line, together with the calculation in the dipole approximation, red dashed line}\label{f3}\end{figure}

%\begin{figure}[h]\center\vskip 15pt\includegraphics[scale=0.8,width=7.8cm,angle=-90]{Fig7.pdf} {All order proton calculation, full black line, plus the calculations done for a neutron of separation energy $S_n=0.137$ MeV and $S_n=0.6$ MeV. In the latter case we show the all order result and the results corresponding to the dipole approximation. }\label{f7}\end{figure}

\begin{figure}[h]
\center
\vskip 15pt
\includegraphics[scale=0.6,width=8cm,angle=0]{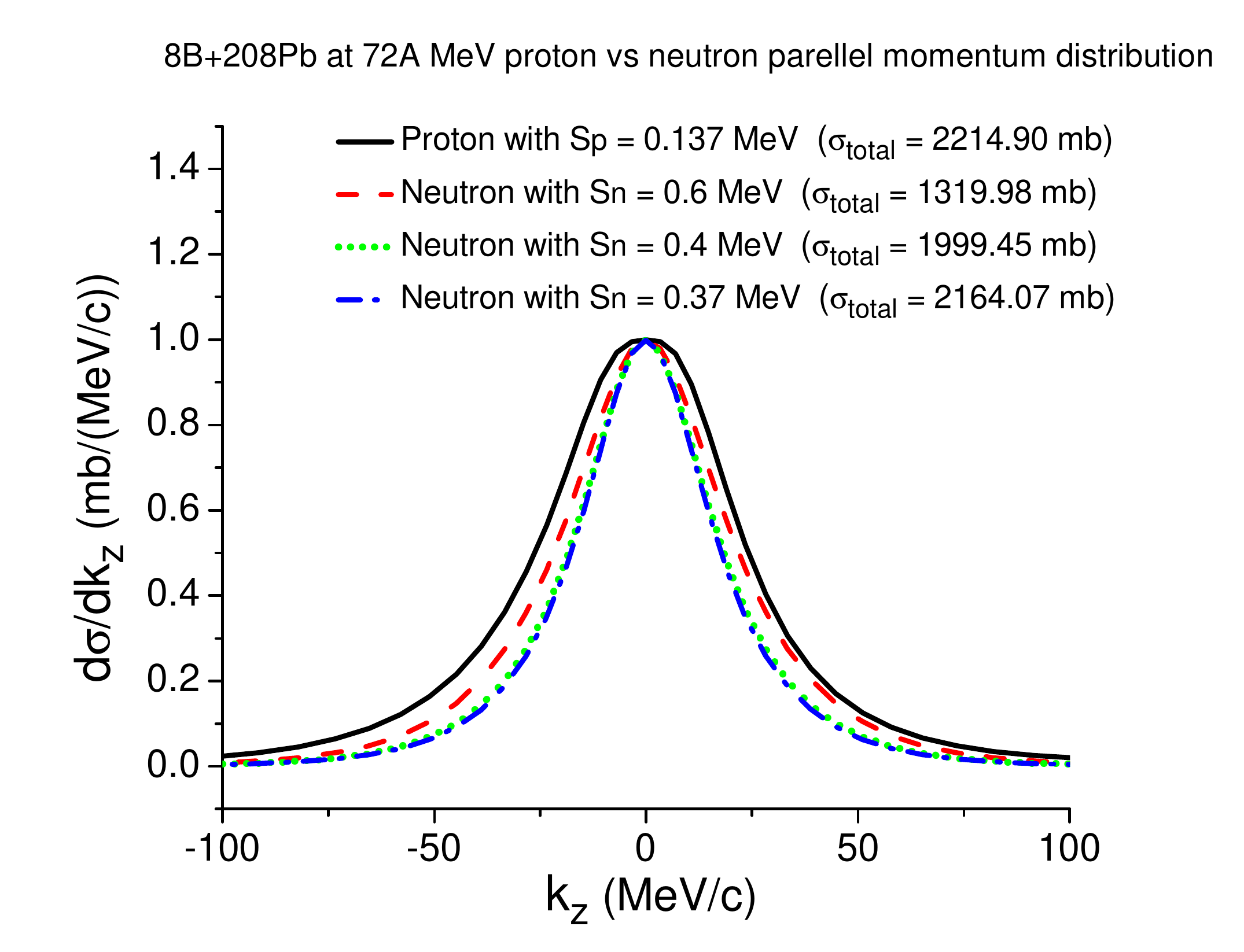}
\caption {(Color online). Differential cross sections for neutrons of different separation energies and proton breakup cross section, all normalized to one. The integrated values are given in the legend.}\label{f8}\end{figure}

%\begin{figure}\center\vskip 15pt\includegraphics[scale=06,width=7cm,angle=-90]{Fig9.pdf}\caption {Stripping part of the nuclear breakup for a proton and a neutron of different separation energies as indicated.}\label{f9}\end{figure}

%\begin{figure}\center\vskip 15pt\includegraphics[scale=0.2,width=7cm,angle=-90]{Fig10.pdf}\caption {Diffraction part of the nuclear breakup for a proton and a neutron of different separation energies as indicated.}\label{f10}\end{figure}

%\begin{figure}\center\vskip 15pt\includegraphics[scale=0.2,width=7cm,angle=-90]{Fig11.pdf}\caption {Direct and recoil terms in Coulomb breakup for the m=0 component of the  $^8$B wave function. }\label{f11}\end{figure}

 In Fig.\ref{f4} we show  separately the effect of the recoil, Eq. (\ref{grec}) term, red dashed line while  the green dotted line represents the direct term obtained from $g^{dir}$. The black full line contains both. It is clear that  the interference between the two is mostly destructive and it becomes constructive only at very large impact parameters. This effect is the equivalent in our model of what Liang et al. \cite{lia} call the "shielding effect" of the proton by the core. The direct term alone, being proportional to $\beta_2$ is indeed in absolute value always larger than the recoil term. However it is  the effect of the interference between direct and recoil terms that causes the reduction of the Coulomb breakup in the proton halo case.

%\begin{figure}\center\vskip 15pt\includegraphics[scale=0.2,width=7cm,angle=-90]{Fig12.pdf}\caption {Direct and recoil terms in Coulomb breakup for given m-component of the wave function as indicated.}\label{f12}\end{figure}

%\begin{figure}\center\vskip 15pt\includegraphics[scale=0.2,width=7cm,angle=-90]{Fig13.pdf}\caption {Total effect of direct and recoil terms in proton  Coulomb breakup including interference compared to the same quantities in the case of a neutron, for the m=0 component of the   $^8$B wave function.}\label{f13}\end{figure}

Following Refs.\cite{lia} and \cite{ang04} we have then calculated the Coulomb breakup to all orders (regularized) and in the dipole approximation for a neutron with several different binding energies. Our hypothesis, that one could use  a neutron wave function corresponding to an {\it effective} separation energy  $S_n$ larger than that of the proton, has been checked for various values of  $S_n$ and we have found that for  $S_n=0.6$ MeV the model works quite well. This is  a factor four smaller than the value  $S_n=2.4$ MeV predicted in  \cite{ang04}. Liang et al. in \cite{lia} found indeed that in the description of the breakup of  $^{17}$F the dipole approximation and perturbation theory to first order could be used by introducing  a neutron wave function corresponding to an effective energy smaller by about a factor three than the one of our original model. The reason is that the model of   \cite{ang04} used an effective binding energy given by an intuitive but not very accurate expression. In this work we use instead Eq.(\ref{1abis}) which through the factors $\beta_1$ and $\beta_2$ is more consistent with the Coulomb potential definition. In order to understand the proton vs neutron breakup dynamics we start by showing in Fig.\ref{f6} LHS the single particle wave function ($S_p=0.137$ MeV) for a p$_{3/2}$ proton with respect to a $^7$Be core by the full black line, for a neutron with the same $S_n$ by the red dashed line and finally for a neutron with $S_n=0.6$ MeV by the green dotted line. The RHS is for $^{17}$F wave function.

%\begin{figure}\center\vskip 15pt\includegraphics[scale=0.2,width=7cm,angle=-90]{Fig14.pdf}\caption {Total effect of direct and recoil terms in proton Coulomb breakup including interference compared to the same quantities in the case of a neutron, for given m-component of the wave function as indicated.}\label{f14}\end{figure}

We have made all order proton calculations and   calculations  for a neutron of separation energy $S_n=0.137$ MeV and $S_n=0.6$ MeV. In the latter case we have studied the impact parameter dependence of the momentum distributions obtained to  all order and compared the results to those  corresponding to the dipole approximation. It is interesting to note that as discussed in \cite{ang04}, and in particular in the appendix of that work, the dipole approximation for a neutron of large binding energy reproduces quite well the proton exact calculation and becomes actually identical to it at large impact parameters.

At this stage we remind the reader that in the case of breakup of a neutron both the "small" width of the momentum distribution and large breakup cross sections were due to the small separation energy and corresponding long tail of the wave function. The proton breakup characteristics are instead quite different as we can interfere from Fig.\ref{f8}. It shows the proton breakup parallel momentum distribution from $^8$B on a $^{208}$Pb target compared to three distributions from an hypothetical neutron bound by three possible energies as indicated. The distributions are normalized to one. The total breakup cross sections including nuclear and Coulomb interactions to all orders are given in the legend. The smallest neutron separation energy case provides the  closest  cross section to the proton's one while it is in the largest separation energy case that the neutron distribution width gets closest to the proton distribution width. This is consistent with the fact that it is the tail of  neutron wave function corresponding to S$_n$=0.6MeV in Fig.\ref{f6} which is closest to the proton's wave function tail.

Therefore we understand that as in the neutron case, the large cross sections are associated with small separation energy but the corresponding momentum distributions have large width because the  wave functions do not have an extended tail. The model of a neutron of larger separation energy representing a proton of smaller separation energy works therefore well for nuclear breakup.  In the case of the proton Coulomb breakup the results are not easy to interpret instead because direct and recoil terms interfer mostly destructively (cf. Fig(\ref{f4}). However the interference becomes constructive at the very large impact parameters where the recoil term is negligible. There, a large breakup can be seen as a consequence of the direct term which as Eq.(\ref{dipp}) shows, being proportional in first order to $\beta_2$, will be large if the proton is well displaced from the core.

We have also studied the spectra corresponding to the m=0,1 components of the wave function for the direct and recoil terms in Coulomb breakup  including interference and we compared to the same quantities in the case of a neutron. The m=0 componenent gives asymmetric distributions while the m=1 component gives  symmetric distributions. These results indicate that  small polarization effects can lead to asymmetric spectra in the proton case and that this is due to the large direct breakup term. Neutron spectra do not show this effects.

 \section{Application to $^{17}$F}

\begin{figure}[h]\center\vskip 15pt\includegraphics[scale=1,width=8cm,angle=0]{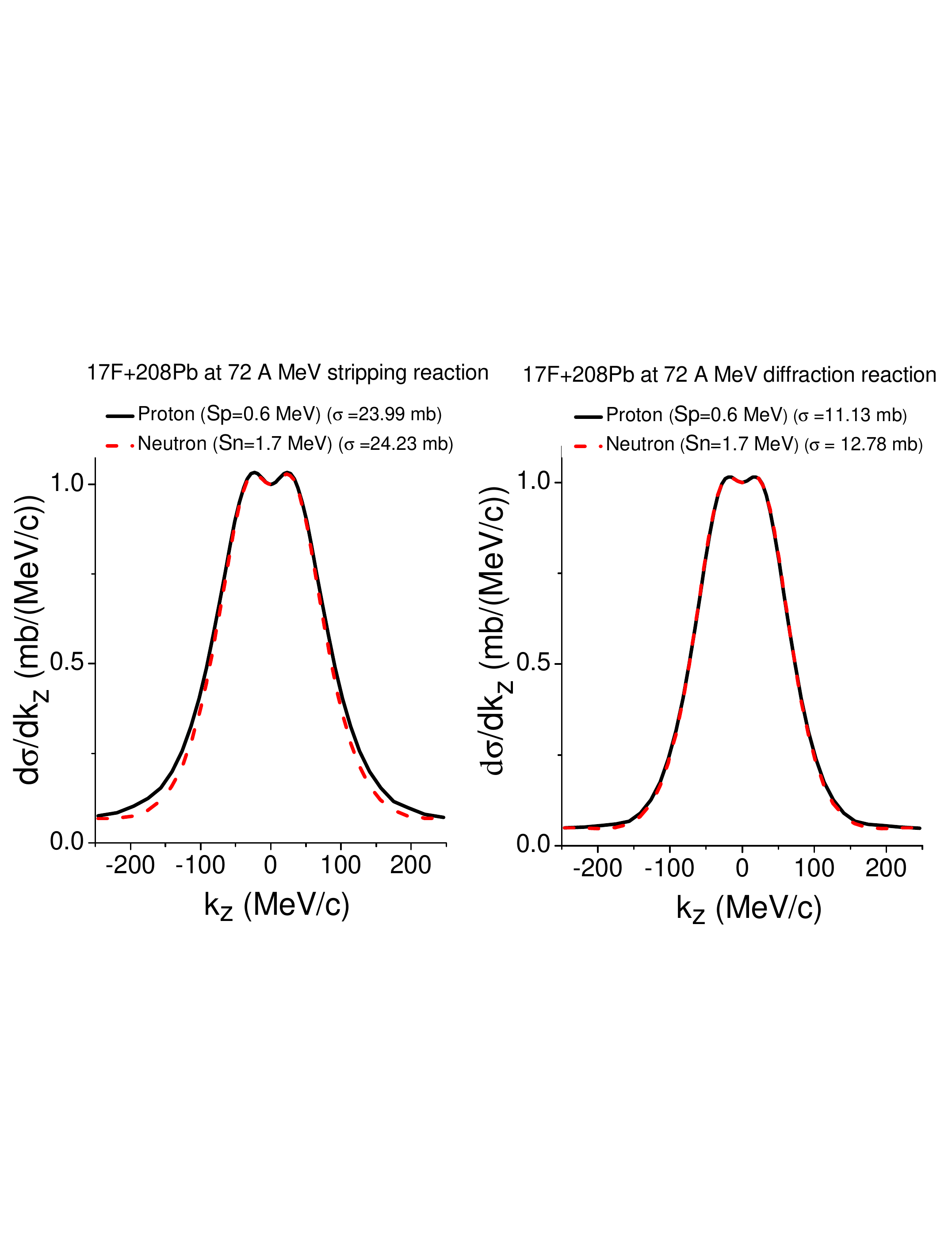}\caption{(Color online). Diffraction and stripping terms of the nuclear breakup of the proton  d$_{5/2}$ ground state in $^{17}$F }\label{f15}\end{figure}

\begin{figure}[h]\center\vskip 15pt\includegraphics[scale=.3,width=6cm,angle=0]{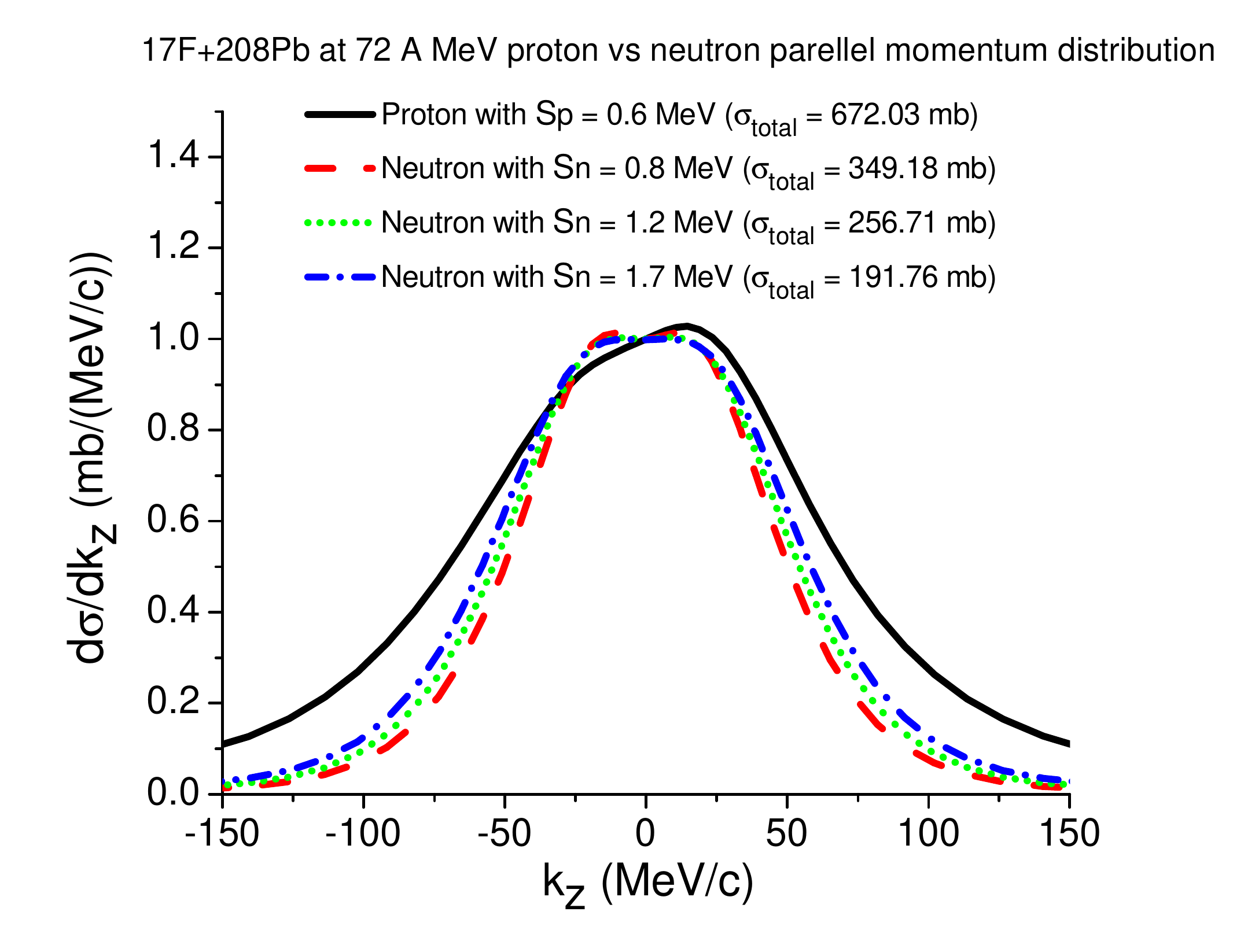}\caption{(Color online). Differential breakup cross sections for neutrons of different separation energies and the proton  (d$_{5/2}$ ground state in $^{17}$F) cross section. All normalized to one. The integrated values are given in the legend.}\label{f17}\end{figure}

\begin{figure}[h]\center\vskip 15pt\includegraphics[scale=1.,width=8cm,angle=0]{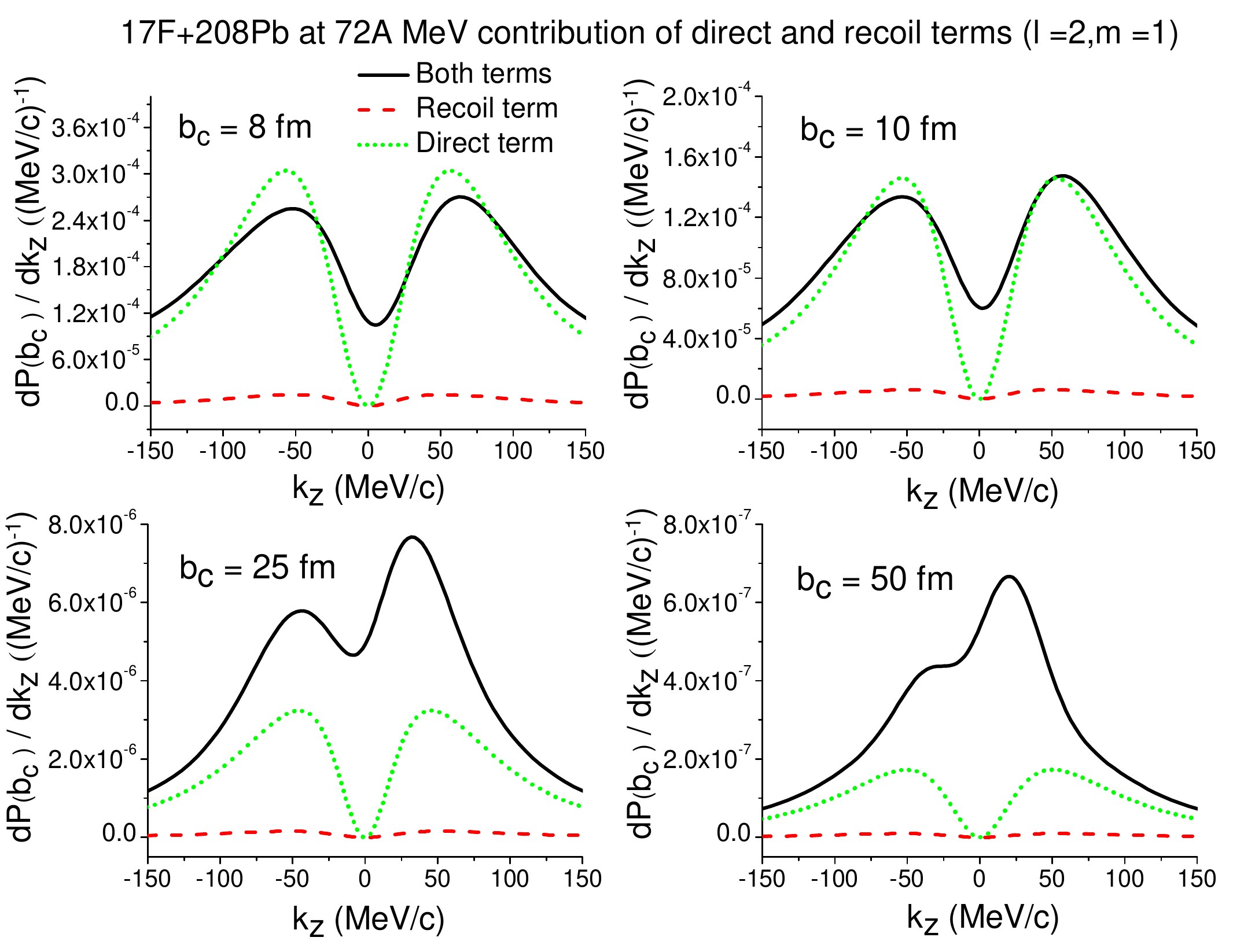}\caption{(Color online). Direct and recoil terms in Coulomb breakup for the m=1 component of the  $^{17}$F wave function.  }\label{f19}\end{figure}

We start this section by suggesting the reader to look at  the proton wave function for the d$_{5/2}$ ground state of $^{17}$F  back in Fig. \ref{f6} . The same figure contains two neutron wave functions for separation energies as indicated that we have used in the calculations shown in the subsequent Figs.\ref{f15}, \ref{f17} and \ref{f19} . %\begin{figure}[h]
%\center
%\vskip 15pt
%\includegraphics[scale=1.,width=8cm,angle=-90]{17Fp_vsNwf.pdf}
%\caption{$^{17}$F d$_{5/2}$ ground state proton wave function and neutron wave functions for binding energies as indicated}
%\label{f15}\end{figure}

By looking at the wave functions in Fig. \ref{f6} and the nuclear breakup diffraction and stripping terms, Fig.\ref{f15} 
we see that the "neutron-like" model works well also for the d$_{5/2}$ ground state in $^{17}$F at 72A.MeV incident energy. The best "model" separation energy here seems to be 1.7MeV. This value is larger than what suggested in \cite{lia} but smaller than the value predicted  in \cite{ang04}. It agrees well with our new effective energy estimate Eq.(\ref{1abis}). 
%\begin{figure}[h]\center\vskip 15pt\includegraphics[scale=1.,width=8cm,angle=-90]{Fig16.pdf}\caption{Direct and recoil terms of the Coulomb breakup at 72 A.MeV of the proton  d$_{5/2}$ ground state in $^{17}$F.}\label{f16}\end{figure}

%\begin{figure}[h]\center\vskip 15pt\includegraphics[scale=1.,width=8cm,angle=-90]{Fig18.pdf}\caption{Direct and recoil terms of the Coulomb breakup at 72 A.MeV of the proton  d$_{5/2}$ ground state in $^{17}$F.}\label{f18}\end{figure}

%\begin{figure}[h]\center\vskip 15pt\includegraphics[scale=1.,width=8cm,angle=-90]{Fig20.pdf}\caption{Direct and recoil terms in Coulomb breakup for the m=0 component of the  $^{17}$F wave function. }\label{f20}\end{figure}

%\begin{figure}
%\center
%\vskip 15pt
%\includegraphics[scale=1.,width=8cm,angle=-90]{PvsN17F10mev.pdf}
%\caption{Direct and recoil terms of the Coulomb breakup of the proton  d$_{5/2}$ ground state in $^{17}$F}
%\label{f19}\end{figure}

To finish this section we show in Fig.\ref{f17} the total cross section distribution for the breakup of the d$_{5/2}$ proton ground state in $^{17}$F at 72 A.MeV, compared to two neutron breakup distributions. What is evident here is a strong asymmetry in the spectrum which instead does not appear in the neutron case. Also in this case the presence of a direct Coulomb breakup term is extremely important because the recoil term is strongly reduced by the centrifugal barrier of the d-state. The immediate consequence is that now the interference between direct and recoil term is mostly constructive.  The asymmetry in the spectrum  originates from the interference of direct and recoil term which is present in the low (m=0,1) components of the initial wave function which dominate the breakup spectrum. Such  effects are  shown by  Fig. \ref{f19}  which contains the probability spectra for the m=1 projection of the initial angular momentum wave function. The m=2 component gives  a symmetrical distribution not shown here.

 \section{Angular Distributions} \label{andis}
  \begin{figure}[h]
\center
\vskip 15pt
\includegraphics[scale=1.6,width=10cm,angle=0]{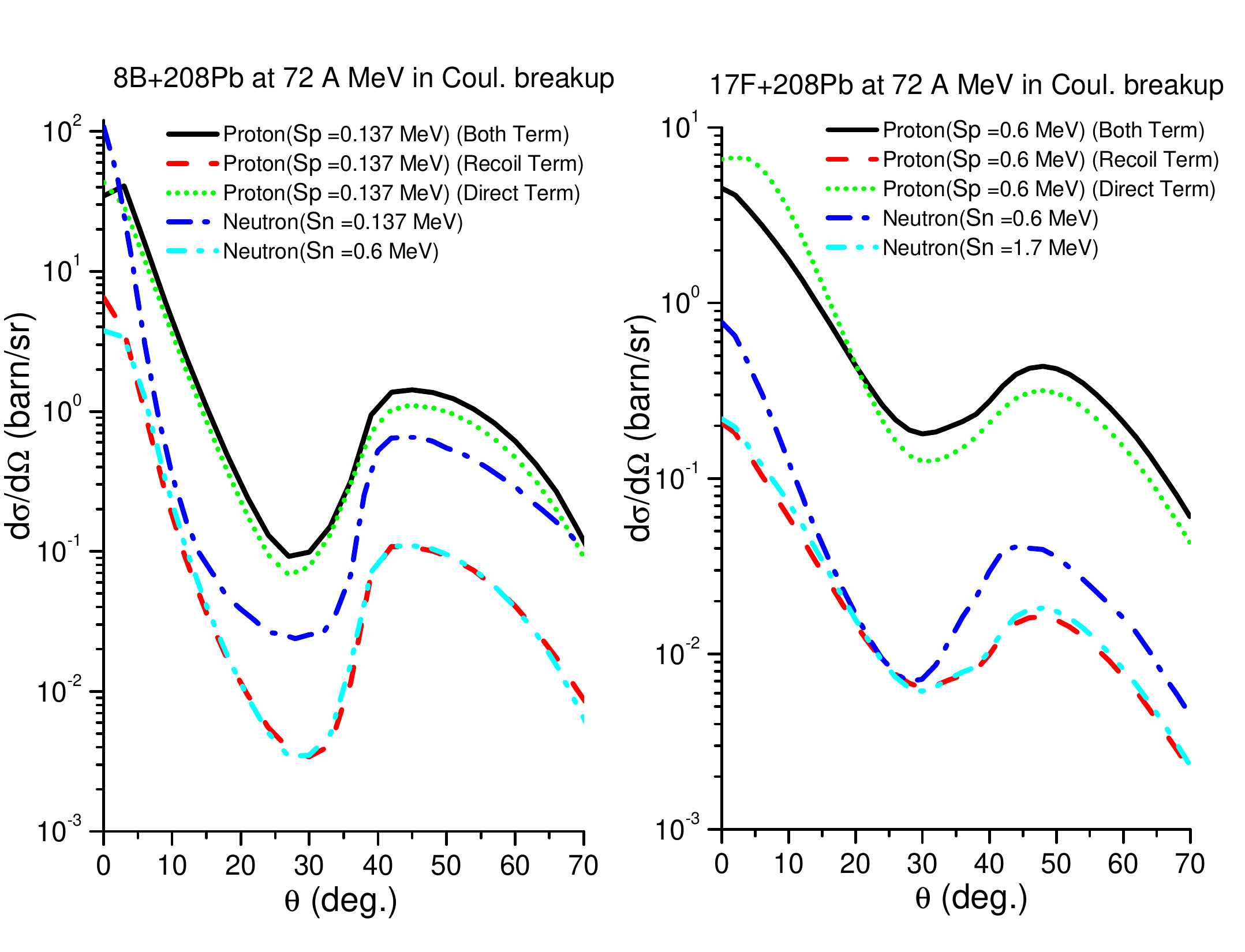}
\caption{(Color online). Proton and neutron angular distributions after breakup. Details are given in the legend.}
\label{f21}\end{figure}

This section is devoted to the discussion of the angular distributions of the breakup protons presented in Fig.\ref{f21}. Here again we show the total result for protons and at the same time the direct and recoil terms. We show also two calculations for neutrons. One for neutron having  the same binding energy as the proton and the other for the effective energy previously obtained. It is very interesting to see that the recoil term of the proton calculation and the  calculation for a neutron with the effective binding energy give very close results. This confirms the interpretation that  the core-target Coulomb effect is  basically a recoil effect for which the difference between a neutron and a proton is that the proton looks "more effectively" bound. On the other hand the direct term gives the most important contribution to the proton breakup, its interference with the recoil term is sometimes constructive, sometimes destructive. This term is not present in the neutron case and it is at the origin of the special dynamical effects of proton breakup
 
 Finally we notice that a neutron of the same binding energy as the proton would give quite a large breakup in the case of the p-state of $^{8}$B, thus suggesting a halo-like behavior for this nucleus, while for $^{17}$F a neutron of the same binding would anyhow give much less breakup at most angles. We can understand this effect by looking at the wave functions in Fig.\ref{f6}. The tail of a neutron wave function for $^{8}$B, being a p-state, is much more pronounced than the tail of  a possible d-neutron wave function in $^{17}$F. Therefore there is no "structural" halo effect in the Fluorine case but the  large proton breakup probability in this case has a dynamical  origin being due to the strong direct proton-target repulsion.  
  \section{Conclusions} \label{concl}
In this paper we have calculated proton  and neutron Coulomb and nuclear breakup from $^{8}$B and $^{17}$F at 72 A.MeV on a $^{208}$Pb target. Our method is an all order formalism based on the eikonal approximation, with a regularized first order Coulomb term. We have compared proton to neutron parallel momentum distributions and angular distributions. The neutron cases have been calculated for an "hypothetical" neutron having the same  separation energy and angular momentum  as the proton in the two projectiles and compared to calculations for a neutron having an {\it effective} binding energy larger than the true proton experimental binding energy.
The effective values take into account the combined effects of the projectile and target Coulomb barrier. We have given an explicit formula to evaluate the effective energy which is in agreement with phenomenological findings.       
 The model has been used to understand the origin of the strong dynamical effects seen in Coulomb breakup data.  Our results clearly show that as far as the nuclear breakup mechanism is concerned
the proton behaves as a neutron of larger separation energy. On the other hand   parallel momentum distributions are   affected by the recoil due to the  neutron-core  Coulomb potential and by the direct repulsion due to proton-target Coulomb potential, in a  complicated way. This happens  both in the calculations, as well as  in the data. There are interference effects between the direct and recoil term which  depend on the impact parameter. Thus even if they can be understood in the theoretical calculation, they would be difficult to disentangle in an experiment. Effects on the core angular distributions are also complicated and will be discussed in a forthcoming publication. However we have finally shown that the proton angular distributions would be  a very good observable to study experimentally in order  to separate the two Coulomb effects. Our results indicate that the proton angular distributions are dominated by the direct term while the recoil term gives  almost an order of magnitude smaller cross section as a function of the proton angle. A neutron of the appropriate {\it effective} binding energy would have exactly the same angular distribution as that given by the recoil term for the proton. On the proton angular distributions  the effects of interference seem to be negligible and we have found some only in the very forward angle region for the $^{17}$F projectile. We have also concluded that a large  breakup cross section for a weakly bound proton  can originate from dynamical effects without corresponding to an unusual structure of the bound state.
 Thus we have clarified the way in which the Coulomb barrier affects weakly bound protons in exotic proton rich nuclei off the stability line. We hope that our results would give some guideline not only in planning  future breakup experiments for spectroscopic and astrophysical studies but more in general in studying the interplay between structure and dynamics of  nuclei with weakly bound protons.

\end{document}